\documentstyle[11pt,aaspp4]{article}

\received{1999 September 1}
\accepted{1999 October 12}
 
\lefthead{Wrobel}
\righthead{Radio Continuum Nucleus in Seyfert Galaxy NGC\,5548}
 
\begin{document}
 
\title{Photometric Variability and Astrometric Stability\\
       of the Radio Continuum Nucleus\\
       in the Seyfert Galaxy NGC\,5548}
 
\author{J.~M. Wrobel}
\affil{National Radio Astronomy Observatory,\\
       P.O. Box O, Socorro, New Mexico 87801\\
       (e-mail: jwrobel@nrao.edu)}

\begin{abstract}
The NRAO VLA and VLBA were used from 1988 November to 1998 June to
monitor the radio continuum counterpart to the optical broad line
region (BLR) in the Seyfert galaxy NGC\,5548.  Photometric and
astrometric observations were obtained at 21 epochs.  The radio
nucleus appeared resolved, so comparisons were limited to observations
spanning 10-60~days and 3-4 yr, and acquired at matched resolutions of
1210~mas $=$ 640~pc (9 VLA observations), 500~mas $=$ 260~pc (9 VLA
observations), or 2.3~mas $=$ 1.2~pc (3 VLBA observations).  The
nucleus is photometrically variable at 8.4~GHz by $33\pm5$\% and
$52\pm5$\% between VLA observations separated by 41~days and 4.1~yr,
respectively.  The 41-day changes are milder ($19\pm5$\%) at 4.9~GHz
and exhibit an inverted spectrum ($\alpha \sim +0.3\pm0.1$,
$S\propto \nu ^{+\alpha}$) from 4.9 to 8.4~GHz.  The nucleus is
astrometrically stable at 8.4~GHz, to an accuracy of 28~mas $=$ 15~pc
between VLA observations separated by 4.1~yr and to an accuracy of
1.8~mas $=$ 0.95~pc between VLBA observations separated by 3.1~yr.
Such photometric variability and astrometric stability is consistent
with a black hole being the ultimate energy source for the BLR, but is
problematic for star cluster models that treat the BLR as a compact
supernova remnant and, for NGC\,5548, require a new supernova event
every 1.7~yr within an effective radius $r_e =$ 42~mas $=$ 22~pc.  A
deep image at 8.4~GHz with resolution 660~mas $=$ 350~pc was obtained
by adding data from quiescent VLA observations.  This image shows
faint bipolar lobes straddling the radio nucleus and spanning
12~arcsec $=$ 6.4~kpc.  These synchrotron-emitting lobes could be
driven by twin jets or a bipolar wind from the Seyfert~1 nucleus.
\end{abstract}


\section{Motivation}

The Seyfert~1 galaxy NGC\,5548 was the target of intensive space- and
ground-based monitoring designed to measure the light travel time
across its broad optical-line-emitting region (\cite{pet93,kor95}, and
references therein).  These reverberation studies imply that the broad
line region (BLR) has a characteristic diameter $\lesssim$~40~light
days, or $\le 64~\mu$as for H$_0=50$~km~s$^{-1}$~Mpc$^{-1}$ and the
galaxy's recession velocity relative to the 3~K background
(5354~km~s$^{-1}$; \cite{dev91}).  Given this small size, is it
possible to constrain models for the ultimate energy source for the
BLR in NGC\,5548?  The common view is that the energy source is a
supermassive black hole fuelled by an accretion disk (e.g.,
\cite{pet97}), and three recent findings support this scenario for
NGC\,5548.  First, analysis of variations in the emission line
profiles of NGC\,5548 suggests that the BLR material moves along
randomly inclined Keplerian orbits defined by a central mass of order
$\sim 7 \times 10^7~M_{\sun}$ (\cite{wan95,wan96,pet99}).  Second, the
central continuum source must irradiate the BLR anisotroptically
(\cite{wan95,wan96}), perhaps because the continuum source is
surrounded by an optically-thick accretion disk (cf.\ Figure~1 of
\cite{goa96}).  Third, NGC\,5548 exhibits very broad and asymmetric
X-ray emission lines whose profiles are well matched by
gravitationally-redshifted and fluorescing lines from accretion disk
material (\cite{nan97}).

Still, the starburst hypothesis of Terlevich et al.\ (1992)
potentially offers an alternative energy source: a compact star
cluster hosts a supernova and the BLR is identified with the compact
supernova remnant formed as the supernova ejecta interacts with dense
circumstellar material.  Although the star cluster hypothesis has yet
to be tested against the three findings listed above for NGC\,5548, it
appears consistent with analysis of the optical continuum variability
and the reverberation results (\cite{are93,ter95}).  Astrometric
monitoring of Seyfert nuclei offers another strong test of of the
starburst hypothesis (\cite{mel97}).  The low supernova rates and
their short radiative lifetimes mean that as the compact star cluster
evolves and new supernovae occur, the BLR should dance within the star
cluster.  In constrast, black hole models for low-power objects like
Seyfert nuclei predict no motion of the nucleus over time, or simple
linear motion if the emission arises from propogating shocks in a jet
and counterjet (\cite{fal96,yi98,yi99}).  This paper imposes
observational constraints on the astrometric stability of the radio
continuum nucleus of NGC\,5548, to help distinguish between star
cluster and black hole models for the energy source of the BLR.
Although the star cluster models do not predict light curves for the
radio continuum emission from a compact supernova remnant, Terlevich
et al.\ (1995) do remark on the relevance of SN\,1988Z to their
models, as being an example of a (radio) supernova expanding into
dense circumstellar material (\cite{sta91,van93}).  Therefore this
paper also examines the photometric variability of the radio continuum
nucleus of NGC\,5548, for comparison with the photometric evolution of
SN\,1988Z at radio wavelengths.

\section{Astrometric Imaging}

Table~1 gives a log of the astrometric imaging observations of
NGC\,5548 with the Very Large Array (VLA, \cite{tho80}) and the Very
Long Baseline Array (VLBA, \cite{nap94}).  Both instruments acquired
data in dual circular polarizations at the center frequencies
tabulated.  For each polarization, the VLA bandwidth was 100~MHz with
3-level sampling and the VLBA bandwidth was 32~MHz with 4-level
sampling.  All VLA and VLBA observations after 1990 assumed a
coordinate equinox of 2000.  Earlier VLA observations assumed an
equinox of 1950 and were transformed to 2000 using a reference date of
1979.9 for the phase calibrator grid used at the VLA prior to 1990
(\cite{wal99}).

\placetable{tab1}

On the VLA, an 8-minute observation of NGC\,5548 was preceded and
followed by an observation of the phase calibrator OQ\,208 (B1404+286)
about $4\arcdeg$ from NGC\,5548 (Figure~1).  VLA observations of
3C\,286 were used to set the amplitude scale to an accuracy of about
3\%.  On the VLBA, a 3-minute observation of NGC\,5548 was preceded
and followed by an observation of the phase, rate, and delay
calibrator J1419+27 about $2\arcdeg$ from NGC\,5548 (Figure~1).  The
position assumed for J1419+27 was $\alpha_{\rm J2000} = 14^h 19^m
59\fs29698$ and $\delta_{\rm J2000} = +27\arcdeg 06\arcmin
25\farcs5540$, with an error in each coordinate of less than 1~mas
(M.\ Eubanks, private communication).  Sources OQ\,208 and J1310+32
were also observed with the VLBA to check the astrometric accuracy and
to align the phases of the baseband channels, respectively.  VLBA
system temperatures and gains were used to set the amplitude scale to
an accuracy of about 5\%, after first correcting for sampler errors.
For both the VLA and VLBA, the position assumed for OQ\,208 was
$\alpha_{\rm J2000} = 14^h 07^m 00\fs3943640$ and $\delta_{\rm J2000}
= +28\arcdeg 27\arcmin 14\farcs69129$ (M.\ Eubanks, private
communication).  These assumed positions for both J1419+27 (B1417+273)
and OQ\,208 agree with those in Ma et al.\ (1998).  The VLA data were
calibrated using the 1994 January 15 release of the NRAO AIPS
software, while the VLBA data were calibrated with the 1999 April 15
release.

\placefigure{fig1}

The AIPS task IMAGR was used to form and deconvolve images of the
Stokes $I\/$ emission from NGC\,5548.  Neither the VLA nor VLBA data
on NGC\,5548 were self-calibrated.  The VLA images were made with a
weighting scheme that achieved high angular resolution without
seriously degrading sensitivity.  Furthermore, VLA imaging tests
showed that the peak source strength increased as angular resolution
decreased, suggesting that the source was slightly resolved.  Thus
only VLA images with matched resolutions were analysed.  The VLBA
images were made with natural weighting to optimize sensitivity and
were restored with a matched elliptical Gaussian beam.  Photometric
and astrometric comparisons will be limited to observations at matched
angular resolutions of 500~mas $=$ 260~pc (Figure~2, 9 VLA
observations), 1210~mas $=$ 640~pc (Figure~3, 9 VLA observations), and
2.3~mas $=$ 1.2~pc (Figure~4, 3 VLBA observations).

\placefigure{fig2}
\placefigure{fig3}
\placefigure{fig4}

Image rms values ($\sigma_I$) appear in Table~1 in units of
microjanskys per beam area ($\mu$Jy~ba$^{-1}$); all rms values match
those expected given the array and observation parameters
(\cite{wro99}).  The image resolutions, expressed either as the FWHM
dimension of a circular Gaussian beam or the FWHM dimensions and
elongation orientation of an elliptical Gaussian beam, also appear in
Table~1.  The tabulated position peaks were derived from quadratic
fits in the image plane.  For the VLA images, these quadratic fits
also yielded the peak flux densities listed in the table, along with
errors that are quadratic sums of a 3\% scale error and $\sigma_I$.
For the VLBA images, the source appeared to be resolved, so total flux
densities were derived by integration over $N \sim 10$ beam areas and
tabulated along with errors that are quadratic sums of a 5\% scale
error and $\sigma_I \sqrt{N}$.  This apparent resolution could be due
either to unmodelled atmospheric and/or station motion effects
(\cite{bea95}), or due to real structure.  That the structure is
similar in the VLBA images suggests the latter but does not prove it,
because some phase-referencing errors can be systematic from
observation to observation.

The table also notes the 1$\sigma_A$ error in each coordinate for the
astrometry of the peak position.  $\sigma_A$ is the quadratic sum of a
term measuring the peak signal-to-noise ratio (\cite{bal75}) and a
term measuring the accuracy of the absolute astrometry.  This latter
term was {\em estimated\/} in the VLA case as follows.  The imaging at
500-mas resolution was done in the same configuration and season as
Patnaik et al.\ (1992) or in the same configuration and
contemporaneous with Browne et al.\ (1998).  Those authors used phase
calibrators $10\arcdeg$ or less from their targets, cycled back to a
calibrator every 6 minutes, and measured a 1-D positional accuracy of
about 10~mas rms.  The present work involves a calibrator $4\arcdeg$
from the target (Figure~1) but observed with a 10-minute cycle time,
so 20~mas was adoted as an estimate for the 1-D accuracy of the
absolute VLA astrometry.  In the VLBA case, the accuracy of the
absolute astrometry was {\em derived\/} from analysis of the
astrometric check source OQ\,208, as referenced to J1419+27.  As
Figure~1 shows, OQ\,208 and J1419+27 are separated by about
$3\arcdeg$, while the separation between J1419+27 and NGC\,5548 is
about $2\arcdeg$.  Thus the position error for OQ\,208 as referenced
to J1419+27, plotted in Figure~5, serves as a cautious upper limit to
the position error for NGC\,5548 as referenced to J1419+27.  Indeed,
residual position errors of order 1~mas and 0.7~mas are expected for
separations of $3\arcdeg$ and $2\arcdeg$, respectively, due to known
deficiencies in the 1980 IAU nutation model adopted at the correlator
(\cite{ma98}).

\placefigure{fig5}

\section{Photometric Analysis}

Photometric monitoring results from Table~1 are plotted in Figure~6.
Each figure panel shows results for a given angular resolution within
100 days centered on the mean observation epoch indicated.  Data at
8.4~GHz are plotted as dark symbols, while data at 4.9~GHz appear as
light symbols.  The figure legend indicates which symbols encode which
matched angular resolutions.  Photometric comparisons should only be
made among data at matched resolutions.  To aid such comparisons the
pairs of panels at the top, middle, and bottom of Figure~6 show data
at matched resolution, with the left columns having the earlier mean
epoch of observation.

\placefigure{fig6}

Figure~6 demonstrates that the radio nucleus of NGC\,5548 is
photometrically variable, on timescales that are both short (comparing
within a panel) and long (comparing across paired panels).
Specifically, the maximum flux densities are higher than minimum flux
densities, by $33\pm5$\% between VLA observations separated by 41~days
(panel [d]) and by $52\pm5$\% between VLA observations separated by
4.1~yr (panel [a] {\em versus} [b]).  Panel (d) also establishes that
the 41-day ``flare'' is milder ($19\pm5$\%) at 4.9~GHz than at
8.4~GHz.  Furthermore, the spectral index $\alpha$ ($S\propto \nu
^{+\alpha}$) is $\alpha \sim -0.2\pm0.1$ prior to the 41-day ``flare''
but $\alpha \sim +0.3\pm0.1$ during the ``flare''.  Thus the ``flare''
is optically thick and milder at the lower frequency, traits typical
for variable active nuclei.  The ``flare'' is too poorly sampled to
derive a meaningful variability timescale.  The flux densities plotted
in Figure~6 correspond to 8.4-GHz powers $P_\nu \sim 2.5-8.3 \times
10^{21}$~W~Hz$^{-1}$, while the associated luminosities are $L_\nu =
\nu P_\nu \sim 2.1-7.0\times 10^{38}$ erg~s$^{-1}$.

The radio sampling in Figure~6 is too coarse for useful comparisons
with concurrent continuum monitoring from the ground or space.
However, some relevant comparison can be made in connection with the
41-day radio variability between 1993 March 15 and April 25.  First,
that radio variability is much slower then the 1-day and 3-day
variability reported for the hard and soft x-rays, respectively,
during the period 1992 December to 1993 January (\cite{don95}).
Second, the 33\% change during the 41-day radio ``flare'' is
comparable to the 40\% changes documented for the continuum at $5100
\AA$ during the period 1993 March 19 to May 27 (\cite{kor95}).

\subsection{Implications for Black Hole Models}

Radio continuum emission can arise from (1) an advection-dominated
accretion flow (ADAF) onto a black hole (\cite{yi98}), (2) jets
emerging from an ADAF (\cite{yi99}), or (3) jets emerging from a
standard accretion disk which itself emits no radio photons
(\cite{fal96}).  Comparisions with these models will assume a black
hole mass $\sim 7 \times 10^7~M_{\sun}$ (\cite{pet99}) and will try to
match the spectral indices $\alpha$, luminosities $L_\nu$, and
photometric variability observed for the radio nucleus of NGC\,5548,
plus be consistent with VLBA emission being predominantly unresolved.

The radio continuum from an ADAF is synchrotron emission fron an X-ray
emitting and optically-thin plasma (\cite{yi98,yi99}), and is
characterized by three key traits.  First, Equation~1 of Yi \& Boughn
(1999) predicts that the radio power should have an inverted spectral
index $\alpha = +0.2$, consistent with the index measured in the
``flaring'' state ($\alpha \sim +0.3\pm0.1$) but not in the the
quiescent state ($\alpha \sim -0.2\pm0.1$).  Second, Equation~2.7 of
Yi \& Boughn (1998) implies that the 8.4-GHz photons should originate
about 6 Schwarzschild radii from the black hole, placing them at radii
less than $0.1~\mu$as, which is consistent with the VLBA emission
being predominantly unresolved.  Finally, Equations~1 and 3 of Yi \&
Boughn (1999) predict an 8.4-GHz luminosity $L_\nu = \nu P_\nu \sim
7\times 10^{36}$ erg~s$^{-1}$.  This is about $30-100$ times weaker
than observed and argues for an alternative source of radio photons.
Radio emission from jets is a plausible alternative, but before
turning to that topic the existing evidence for outflow from the
nucleus of NGC\,5548 will be described.

Figure~7 displays a VLA image at 8.4~GHz with resolution 660~mas $=$
350~pc and rms sensitivity $\sigma_I = 12~\mu$Jy~ba$^{-1}$.  This deep
image was obtained by adding the 1988-1989 VLA data lacking evidence
for photometric variability (see panels [a] and [c] in Figure~6) and
two further 1989 observations at lower resolution.  Areal integration
over the figure yields a total flux density of $8.5\pm0.3$~mJy,
corresponding to a total power at 8.4~GHz of $1.2 \times
10^{21}$~W~Hz$^{-1}$.  Figure~7 shows that bipolar lobes straddle the
radio nucleus and span 12~arcsec $=$ 6.4~kpc.  The lobes are very
faint at 8.4~GHz, with peak intensities $< 160~\mu$Jy~ba$^{-1}$.  The
lobe morphologies confirm previous hints from images at lower
resolution (Figure~3, \cite{wil82,bau93,kuk95,nag99}) that the
southern lobe is amorphous and the northern lobe is edge-brightened,
possibly due to a bubble-like structure in 3-D.

\placefigure{fig7}

These synchrotron-emitting lobes could be driven by a bipolar wind
emanating from the nucleus of NGC\,5548, as has been suggested for
NGC\,3079 (\cite{baa95,tro98}).  Models for the absorption of the
X-ray and ultraviolet continuum from NGC\,5548 do imply outflowing
ionized gas on subparsec scales (\cite{mat95}).  The shape of the
subparsec outflow (radial {\em versus} bipolar {\em versus}
cylindrical) is not constrained, but its kinetic luminosity of
$10^{43}$~erg~s$^{-1}$ is four orders of magnitude higher than the
photon luminosities of the radio lobes (\cite{wil82}).  If the
subparsec outflow contains relativistic plasma, then Figure~4 could be
used to trace the shape of the flow on parsec scales provided the
reality of the apparent VLBA structure was verified (see Section~2).
It is worth noting that weak synchrotron emission from relativistic
plasma on scales 100-500~pc could be responsible for the source
resolution effects, discussed in Section~2, that prompted the use of
matched angular resolutions.

Suppose the synchrotron luminosity of the jets on scales $\lesssim
500$~pc is powered by a rotating black hole accreting from a
magnetized plasma.  Then the jet luminosity can be estimated from the
Blandford-Znajek process given values for the spin paramter $\bar{a}
\le 1$ of the black hole, the efficiency $\epsilon_{\rm jet} \le 1$ of
the jet radio emission, and the mass accretion rate $\dot{m}$,
normalized to the Eddington rate (\cite{yi99}).  Adopting $\dot{m}$
from an ADAF analysis of NGC\,5548 leads to radio jet luminosities of
$L_\nu \sim 8.4\times 10^{43} \bar{a}^2 \epsilon_{\rm jet}$
erg~s$^{-1}$.  This process could easily match the required radio jet
luminosites on scales $\lesssim 500$~pc, but the predictions are
woefully imprecise compared to the predictions from the ADAF models.
This process could also in principal match the kinetic luminosity of
$10^{43}$~erg~s$^{-1}$ for the subparsec outflow (\cite{mat95}).
Alternatively, in the jet-disk symbiosis model of Falcke \& Biermann
(1996), the radio jet luminosity and the (standard) accretion disk
luminosity are correlated.  Gauging the latter for NGC\,5548 by the
nuclear bolometric luminosity of $5\times 10^{44}$~erg~s$^{-1}$
(\cite{mat95}), then the radio jet luminosity is as expected from that
jet/disk correlation.  Moreover, the spectral index of the quiescent
radio emission is as expected for an inhomogeneous jet model, while
the inverted spectrum during the 41-day ``flare'' could arise from a
brightening in the optically-thick base of the jet.

\subsection{Implications for Star Cluster Models}

Terlevich et al.\ (1995) note the relevance of SN\,1988Z to their
models, as being an example of a (radio) supernova expanding into
dense circumstellar material (\cite{sta91}).  It is thus of interest
to compare the radio properties of this most distant known radio
supernova (\cite{van93}), with those of the radio nucleus in
NGC\,5548.  Van Dyk et al.\ (1993) conclude that SN\,1988Z is a rare,
peculiar Type II supernova, which represents the endpoint of stellar
evolution for a massive progenitor ($\sim 20-30~M_{\sun}$) that
underwent a period of high mass loss ($\gtrsim
10^{-4}~M_{\sun}$~yr$^{-1}$) before explosion.  SN\,1988Z achieved a
maximum power at 8.4~GHz of about $2 \times 10^{21}$~W~Hz$^{-1}$,
similar to the minimum power recorded in Figure~6 for the radio
nucleus of NGC\,5548.  Van Dyk et al.\ (1993) present models
quantifying the photometric evoluton of SN\,1988Z at 8 and 5~GHz.
(While the evolution appears to be smooth, the defining light curves
are not well sampled in time.)  Their models are plotted in Figure~8
with an arbitrary epoch shift for comparison with the VLA data from
panel (d) of Figure~6.  The radio supernova models appear able to
account for the photometric variability of NGC\,5548 observed on time
scales of years.  Those models cannot, however, account for the
variability observed on time scales of tens of days.

\placefigure{fig8}

\section{Astrometric Analysis}

All astrometric monitoring results from Table~1 are plotted in
Figure~9.  Each panel shows data at the matched angular resolution
indicated, with the panels ordered as in Figure~6 to aid astrometric
comparisons among data at matched resolutions.  The origin for each
panel is taken to be the position of the nucleus measured from the
VLBA image at epoch 1995.337.  Data at 8.4~GHz are plotted as dark
symbols, while data at 4.9~GHz appear as light symbols.  The symbol
shapes encode the matched angular resolution and have the same meaning
as in Figure~6.

\placefigure{fig9}

Figure~9 implies that the radio nucleus of NGC\,5548 is
astrometrically stable, on timescales that are both short (comparing
within a panel) and long (comparing across paired panels).  Panels (a)
and (b) demonstrate that the nucleus is astrometrically stable at
8.4~GHz, to an accuracy of $20\sqrt{2}$~mas $=$ 28~mas $=$ 15~pc
between VLA observations spanning both tens of days and 4.1~yr.
Panels (c) and (d) verify this stability but with a poorer accuracy of
$25\sqrt{2}$~mas $=$ 35~mas $=$ 19~pc.  The nucleus is as
astrometrically stable at 4.9~GHz as it is at 8.4~GHz, based on
comparisons within panels (b) and (d).  Finally, panels (e) and (f)
imply that the nucleus is astrometrically stable at 8.4~GHz, to an
accuracy of $1.3\sqrt{2}$~mas $=$ 1.8~mas $=$ 0.95~pc between VLBA
observations spanning 13 days and 3.1~yr.

\subsection{Implications for Black Hole Models}

The observed astrometric stability is consistent with any models based
on black holes, including ADAF and jet models, but an ADAF model was
ruled out photometrically in Section~3.1.  If the reality of the
apparent VLBA structure in Figure~4 could be verified, then those
images could be used to constrain the evolution of the lengths of, or
features within, any jets or counterjets.  Furthermore, if a VLBA
image was acquired during a radio flare, then the peak in that image
would arise from the optically-thick base of the jet and would be
expected to be located closer to the black hole.

\subsection{Implications for Star Cluster Models}

Suppose that the blue luminosity from the Seyfert ``nucleus'' of the
NGC\,5548 arises from a compact star cluster.  Then the absolute
magnitude of the cluster can be equated to the historical minimum of
$M_B = -20.6$~mag recorded for the Seyfert nucleus (\cite{are93}).  A
star cluster with that magnitude will be characterized by an optical
supernova rate of 0.6~yr$^{-1}$ (\cite{are94}) and an effective radius
$r_e = 22$~pc $=$ 42~mas (\cite{you76,mel97}).  Each new optical
supernova event should yield a radio counterpart if SN\,1988Z is
indeed as relevant to the starburst models as claimed by Terlevich et
al.\ (1995).

A circle with the effective radius $r_e$ of the star cluster is shown
in panels (a)-(d) in Figure~9, with the circle arbitrarily centered at
the origin.  Figure~9 implies that radio nucleus of NGC\,5548 does not
move with the characteristic step size $r_e$ between supernova events,
as predicted by Melnick et al.\ (1997).  The strongest constraints
follow from the VLBA-VLBA astrometric stability at the 1-pc level for
observations separated by 3.1~yr.  However, in the star cluster model
it is the massive stars which are the supernova progenitors, so if the
initial mass function is biased toward massive stars and/or mass
segregation is occuring, then the radio supernova would occur within a
region of radius much smaller than $r_e = 22$~pc $=$ 42~mas
(\cite{mel97}).

The optical supernova rate of 0.6~yr$^{-1}$ for NGC\,5548 is
moderately high, leading to a new supernova event about every 1.7~yr.
This motivates a deep search for secondary radio sources at the
matched VLBA resolution.  Data from VLBA epochs 1998.445 and 1998.481
were combined to yield an image with rms $\sigma_I =
41~\mu$Jy~ba$^{-1}$, corresponding to about 100 times the strength of
Casseopia\,A were it placed at the distance of NGC\,5548
(\cite{baa77}).  No secondary sources stronger than $5\sigma_I$, or
about 200 times Casseopia\,A, were found in about 7600 VLBA beam areas
within an annular region, centered as for Figure~9 and with inner
radius 10~mas $=$ 5.3~pc and outer radius 84~mas $=$ 45~pc $= 2r_e$.
This outer radius is well within the limitations to the field of view
given the VLBA observing strategies (\cite{wro95}).  The absence of
secondary VLBA sources, in conjuction with the VLBA constraints on the
astrometric stability of the radio source, requires the presence of no
more than one radio supernova during a 3.1-yr interval.  This
condition is improbable if the detected VLBA source really is a radio
supernova whose slow decline, based on the VLBA photometry, is shared
by other radio supernovae.  Thus either the radio supernova rate is
lower than 0.6~yr$^{-1}$ and/or, as mentioned above, the radio
supernovae occur within a region of radius much smaller than $r_e =
22$~pc $=$ 42~mas.

\section{Conclusions and Future}

The VLA and VLBA were used to monitor the radio continuum counterpart
to the optical broad line region (BLR) in the Seyfert galaxy
NGC\,5548.  The nucleus is photometrically variable at 8.4~GHz by
$33\pm5$\% and $52\pm5$\% between VLA observations separated by
41~days and 4.1~yr, respectively.  The 41-day changes are less
pronounced ($19\pm5$\%) at 4.9~GHz and exhibit an inverted spectrum
($\alpha \sim +0.3\pm0.1$, $S\propto \nu ^{+\alpha}$) from 4.9 to
8.4~GHz.  The nucleus is astrometrically stable at 8.4~GHz, to an
accuracy of 15~pc between VLA observations separated by 4.1~yr and to
an accuracy of 0.95~pc between VLBA observations separated by 3.1~yr.

The photometric variability and astrometric stability of the radio
nucleus is not consistent with star cluster models that treat the BLR
as a compact supernova remnant and, for NGC\,5548, required a new
supernova event like SN\,1988Z every 1.7~yr within an effective radius
$r_e = 22$~pc.  The radio supernova models for SN\,1988Z cannot
account for the VLA variability observed on time scales of tens of
days.  The combined VLBA evidence for astrometric stability, no
secondary sources, and slow photometric evolution of the detected
source mean that the radio supernova rate is lower than predicted
and/or the radio supernovae occur within a region much smaller than
predicted.  Additional VLBA monitoring would thus be benefical, to
provide further astrometric and photometric constraints on the
detected radio source and to continue searches for secondary sources.

The photometric variability and astrometric stability of the radio
nucleus in NGC\,5548 is consistent with models of jets on scales
$\lesssim 500$~pc emerging from a black hole.  If the subparsec
outflow inferred from X-ray/UV absorber models contains relativistic
plasma, then the VLBA images could be used to trace the shape of the
flow on parsec scales.  However, the reality of the apparent VLBA
structure must first be verified.  Tracing flow kinematics to 100~pc
should also be feasible through an HST study of the narrow optical
emission lines, because those lines are so dominated by gas within
70~pc $=$ 130~mas from the nucleus (\cite{kra98} and references
therein).

A deep image at 8.4~GHz with a resolution 350~pc was obtained by
adding data from quiescent VLA observations.  This image shows faint
bipolar lobes straddling the radio nucleus and spanning 6.4~kpc.  The
northern lobe is edge-brightened, suggesting a bubble-like structure
in 3-D.  These lobes are plausibly driven by twin jets or a bipolar
wind from the Seyfert~1 nucleus.  The kinematics of the flow could be
followed to kiloparsec scales if neutral or molecular gas was detected
in absorption against the radio continuum lobes.  Sensitive
polarimetry of the radio lobes could also constrain the properties of
an enveloping magnetoionic medium and the topology of the magnetic
field in a synchrotron-emitting bubble.

\acknowledgments The author thanks Dr.\ M.\ Eubanks for providing VLBI
source positions; thanks Drs.\ J.\ Conway, R.\ Terlevich, J.\ Ulvestad
and C.\ Walker for discussions; and acknowledges support from a SERC
Visiting Fellowship held at Nuffield Radio Astronomy Laboratories,
England.  NRAO is a facility of the National Science Foundation
operated under cooperative agreement by Associated Universities, Inc.

\clearpage

\figcaption[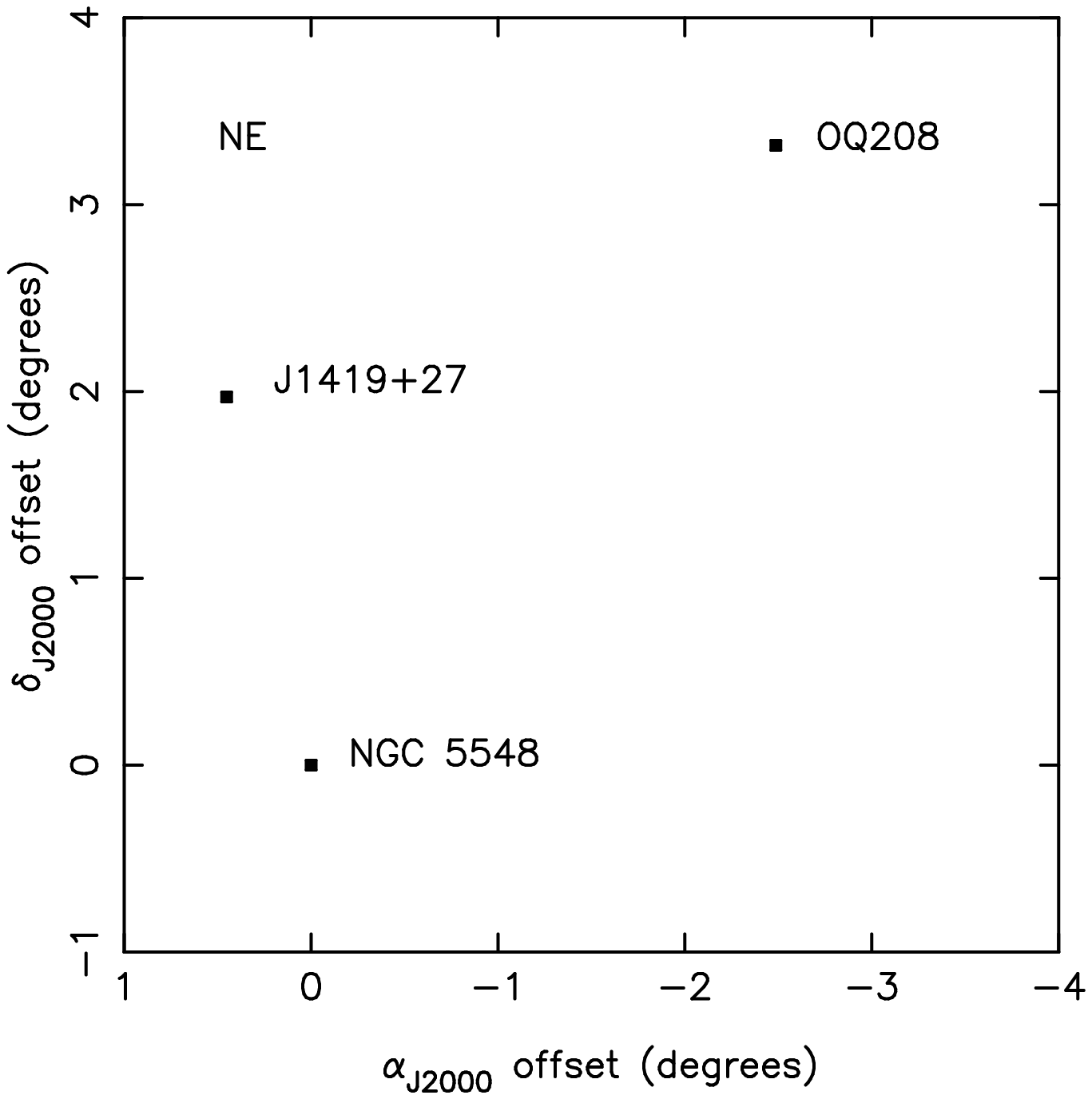]{Phase reference sources near NGC\,5548.
\label{fig1}}

\figcaption[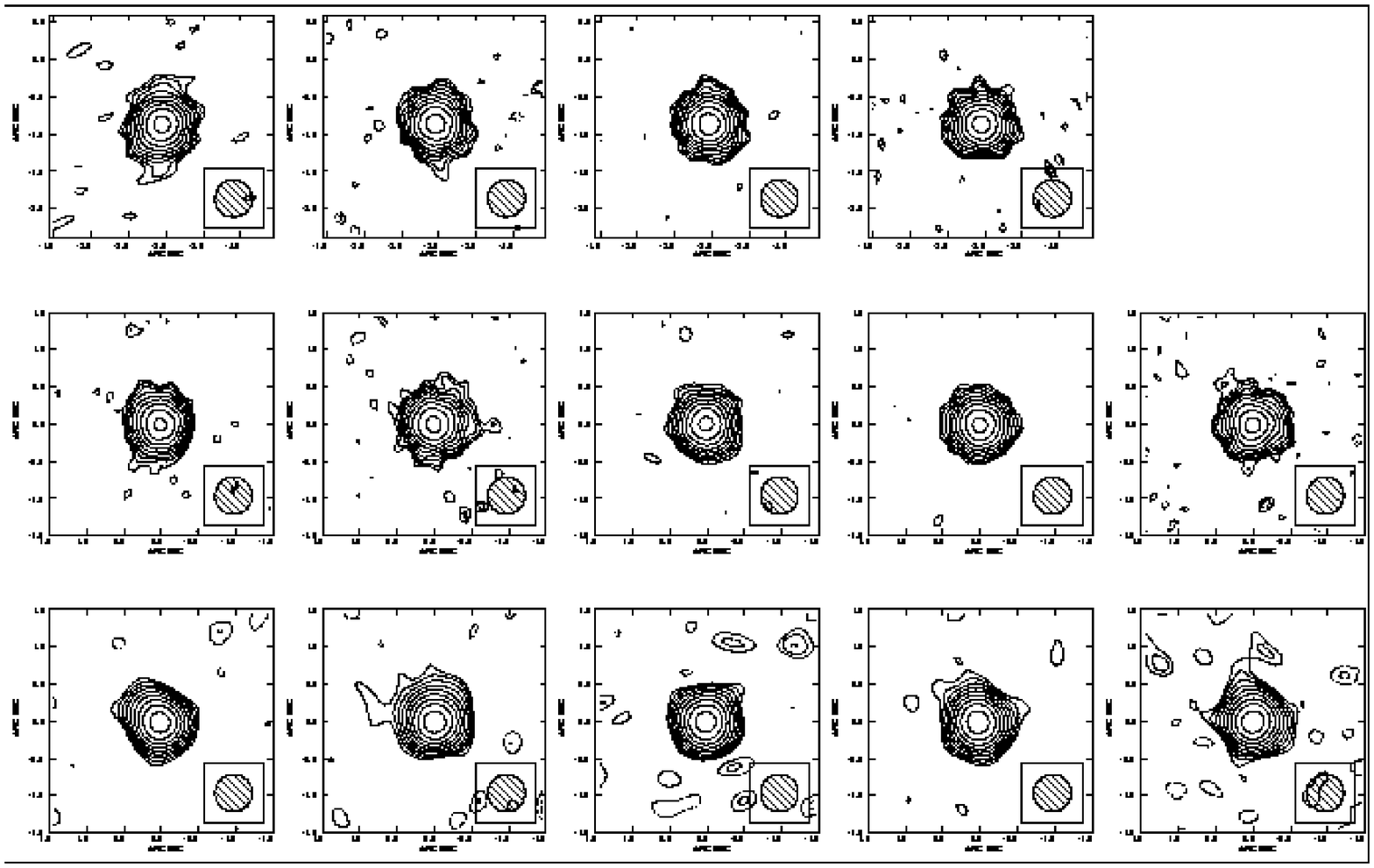]{Montage of VLA images of Stokes $I\/$ emission from
NGC\,5548 at a matched resolution of 500~mas $=$ 260~pc (hatched
circle).  Contours are at intervals of $2^{1\over2}$ times the bottom
contour level ($BCL\/$).  Negative contours are dashed and positive
ones are solid.  Image peaks appear in Table~1.  {\em Upper row:}
Frequency 8.4~GHz. $BCL = 100~\mu$Jy~ba$^{-1}$.  Epochs left to right
are 1988.913, 1988.989, 1989.036, and 1989.077.  {\em Middle row:}
Frequency 8.4~GHz. $BCL = 150~\mu$Jy~ba$^{-1}$.  Epochs left to right
are 1992.814, 1992.858, 1992.891, 992.997, and 1993.025.  {\em Bottom
row:} Frequency 4.9~GHz. $BCL = 150~\mu$Jy~ba$^{-1}$.  Epochs left to
right are as for the middle row.
\label{fig2}}

\figcaption[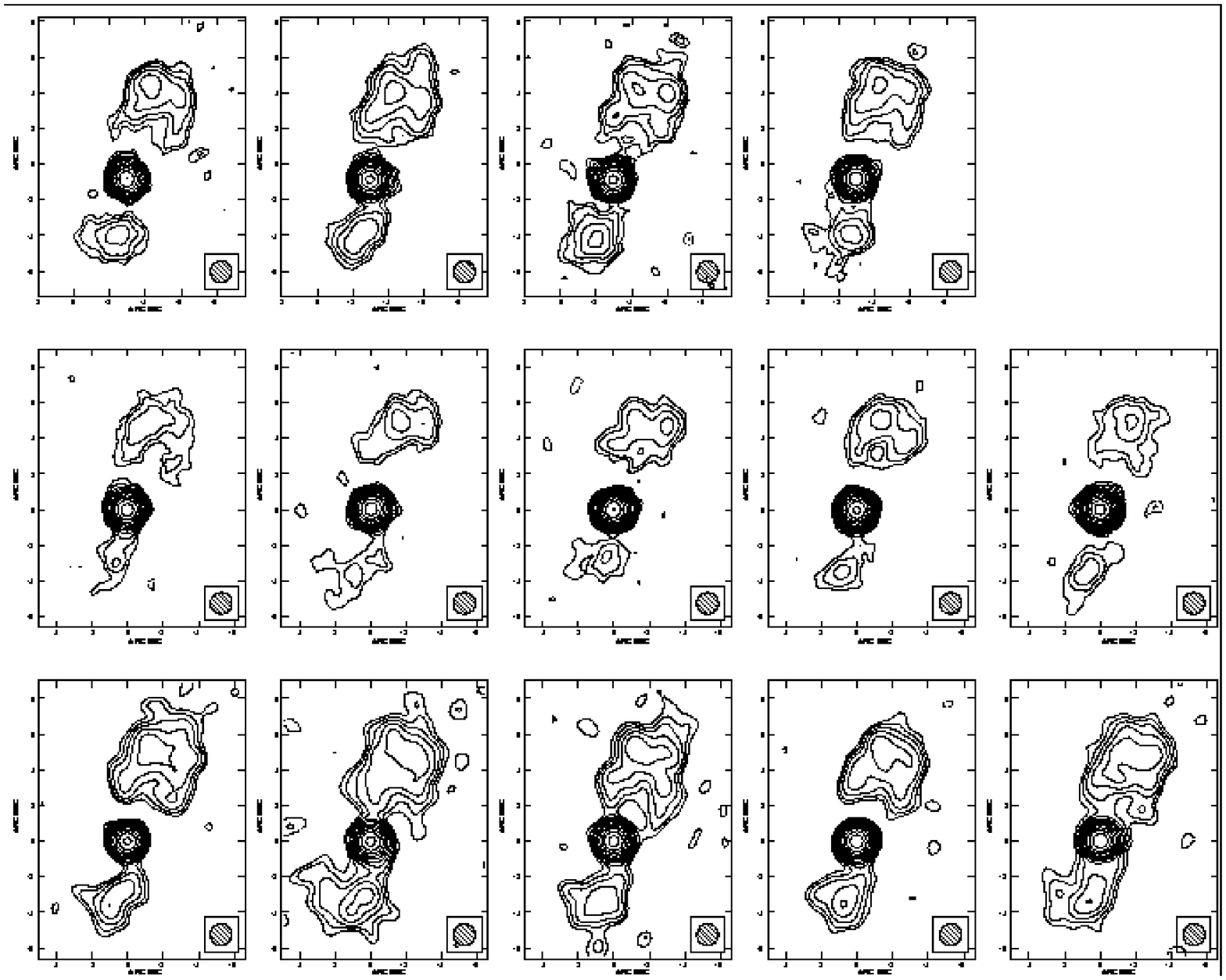]{Montage of VLA images of Stokes $I\/$ emission from
NGC\,5548 at a matched resolution of 1210~mas $=$ 640~pc (hatched
circle).  Contours are at intervals of $2^{1\over2}$ times the bottom
contour level ($BCL\/$).  Negative contours are dashed and positive
ones are solid.  Image peaks appear in Table~1.  {\em Upper row:}
Frequency 8.4~GHz. $BCL = 100~\mu$Jy~ba$^{-1}$.  Epochs left to right
are 1989.184, 1989.230, 1989.279, and 1989.318.  {\em Middle row:}
Frequency 8.4~GHz. $BCL = 150~\mu$Jy~ba$^{-1}$.  Epochs left to right
are 1993.162, 1993.203, 1993.249, 1993.277, and 1993.315.  {\em Bottom
row:} Frequency 4.9~GHz. $BCL = 150~\mu$Jy~ba$^{-1}$.  Epochs left to
right are as for the middle row.
\label{fig3}}

\figcaption[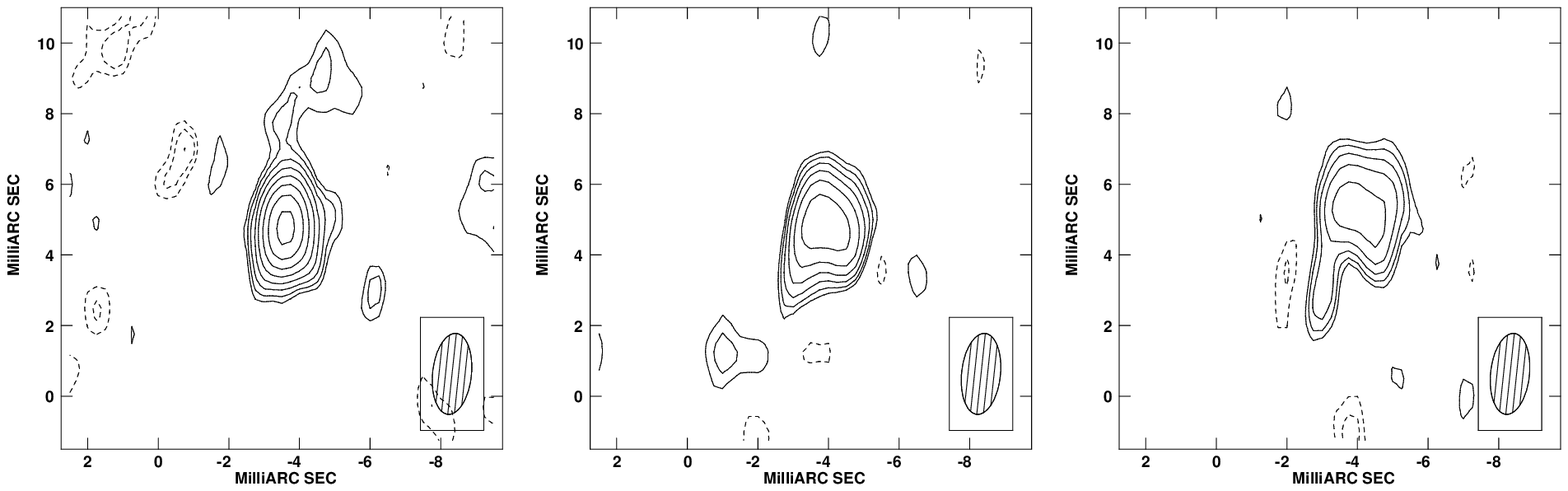]{Montage of VLBA images of Stokes $I\/$ emission
from NGC\,5548 at a frequency of 8.4~GHz and a matched resolution of
2.3~mas $\times$ 1.1~mas $=$ 1.2~pc $\times$ 0.58~pc (hatched
ellipse).  Contours are at intervals of $2^{1\over2}$ times the bottom
contour level of $150~\mu$Jy~ba$^{-1}$.  Negative contours are dashed
and positive ones are solid.  Total flux densities appear in Table~1.
{\em Left:} Peak 1.93~mJy~ba$^{-1}$.  Epoch 1995.337.  {\em Middle:}
Peak 1.09~mJy~ba$^{-1}$.  Epoch 1998.445.  {\em Right:} Peak
0.73~mJy~ba$^{-1}$.  Epoch 1998.481.
\label{fig4}}

\figcaption[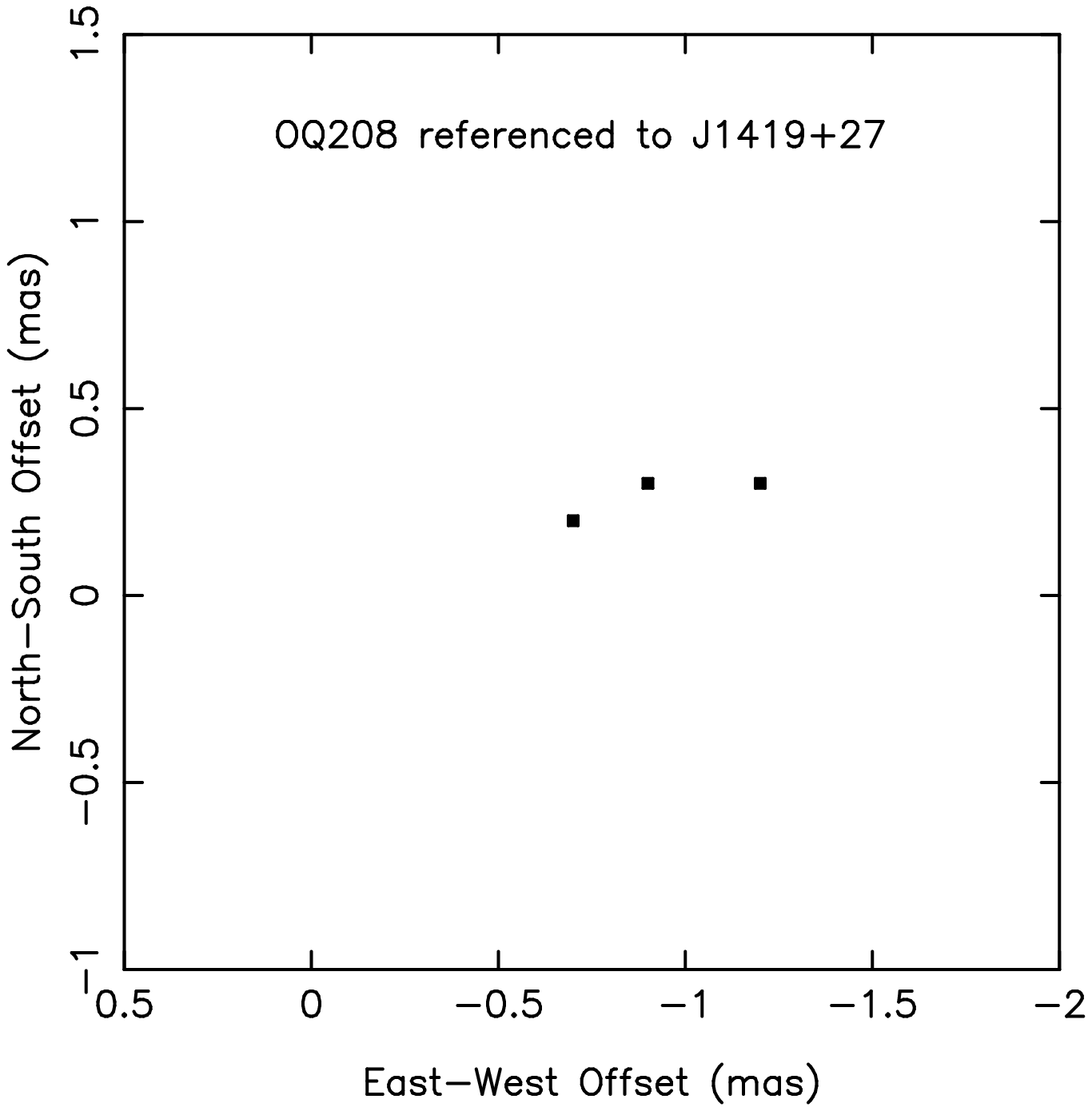]{Observed minus expected VLBA position for OQ\,208
at UT epochs 1995.337, 1998.445, and 1998.481.
\label{fig5}}

\figcaption[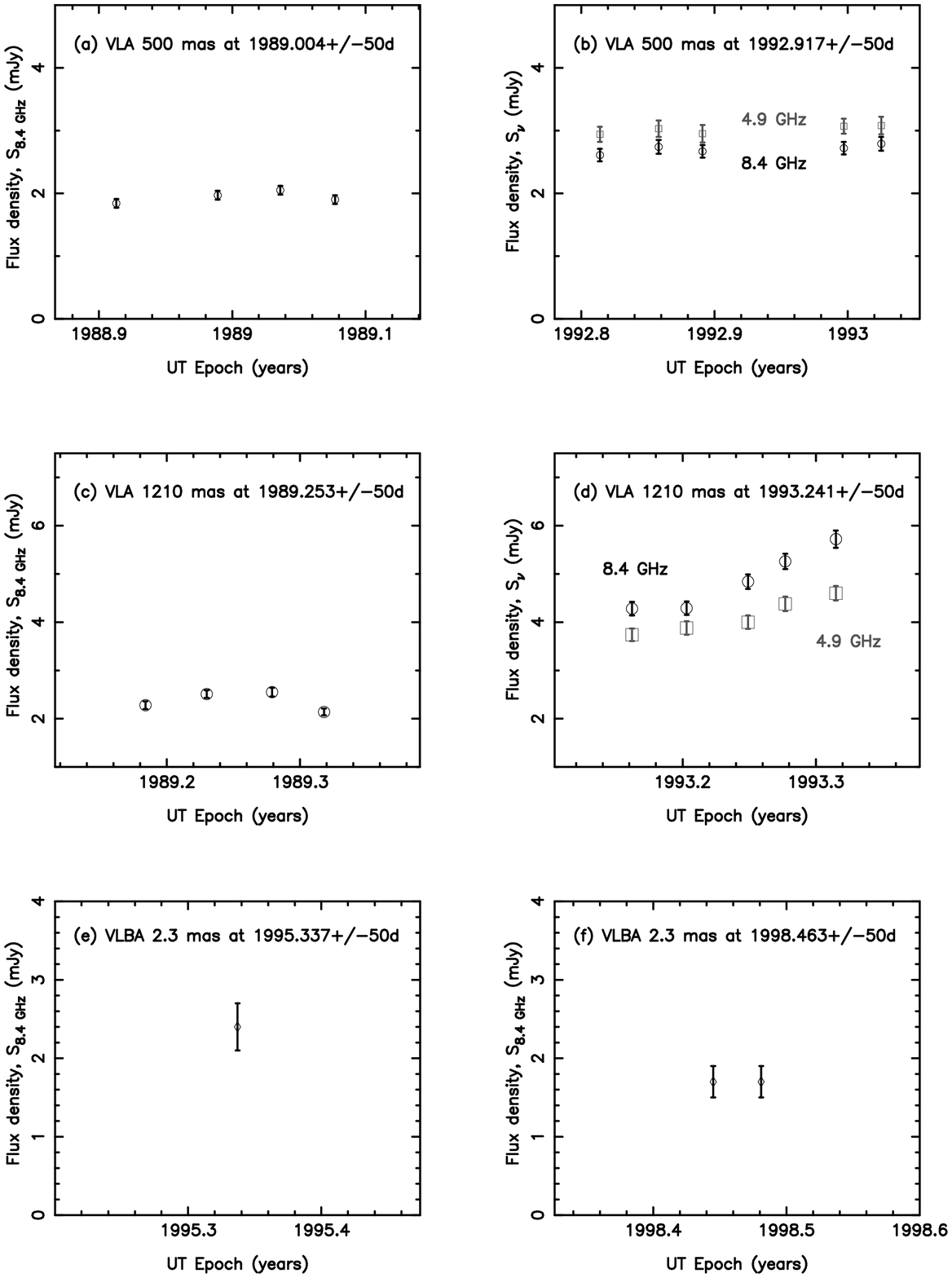]{Photometric monitoring of the radio continuum
nucleus of NGC\,5548 at the matched angular resolutions and mean
epochs indicated.  Data are from Table~1.  Small circles and squares
symbolize peak flux densities at a resolution of 500~mas $=$ 260~pc
(see Figure~2).  Big circles and squares indicate peak flux densities
at a resolution of 1210~mas $=$ 640~pc (see Figure~3).  Small diamonds
correspond to integrated flux densities at a resolution of 2.3~mas $=$
1.2~pc times 1.1~mas $=$ 0.58~pc (see Figure~4).
\label{fig6}}

\figcaption[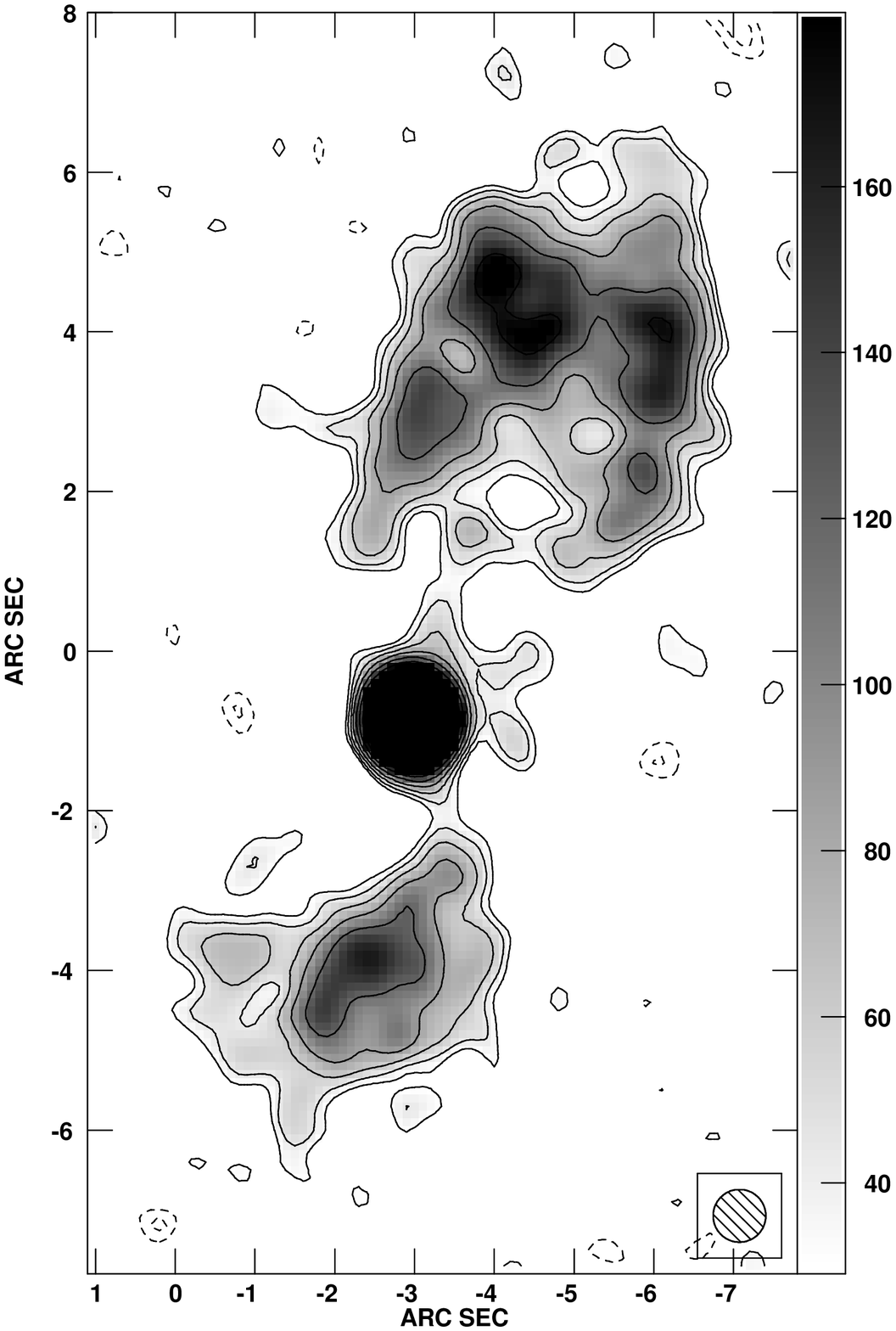]{Deep VLA image of Stokes $I\/$ emission from
NGC\,5548 at a frequency of 8.4~GHz, resolution of 660~mas $=$ 350~pc
(hatched circle), and a mean epoch of 1989.192.  Contour levels are at
intervals of $2^{1\over2}$ times the bottom contour level, which is
$30~\mu$Jy~ba$^{-1}$.  Negative contours are dashed and positive ones
are solid.  Grey scale ranges linearly from 30 to
180$~\mu$Jy~ba$^{-1}$.
\label{fig7}}

\figcaption[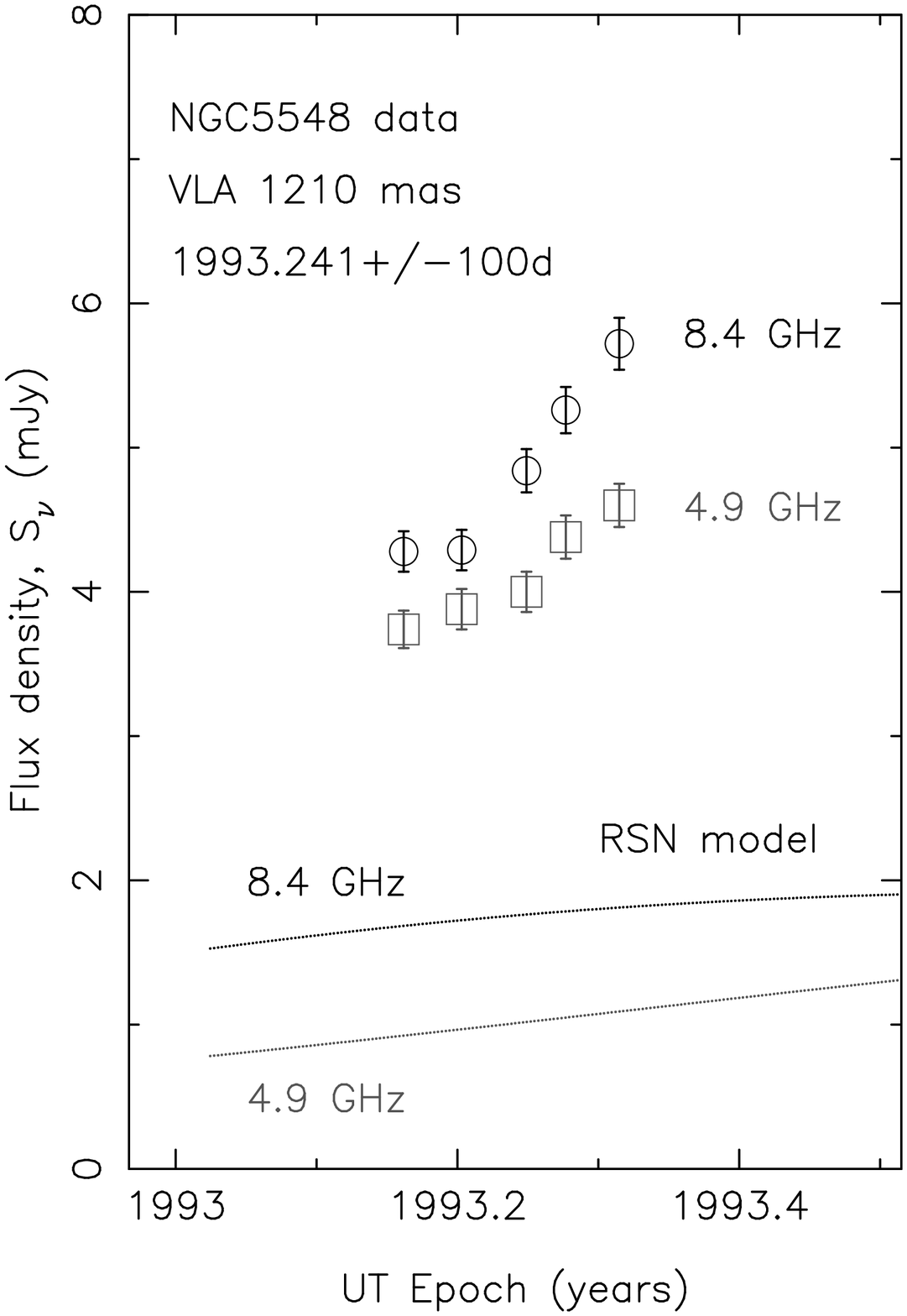]{Photometric monitoring of the radio continuum
nucleus of NGC\,5548 during the 41-day ``flare'', for comparison with
models for the photometric evolution of the radio supernova SN\,1988Z.
\label{fig8}}

\figcaption[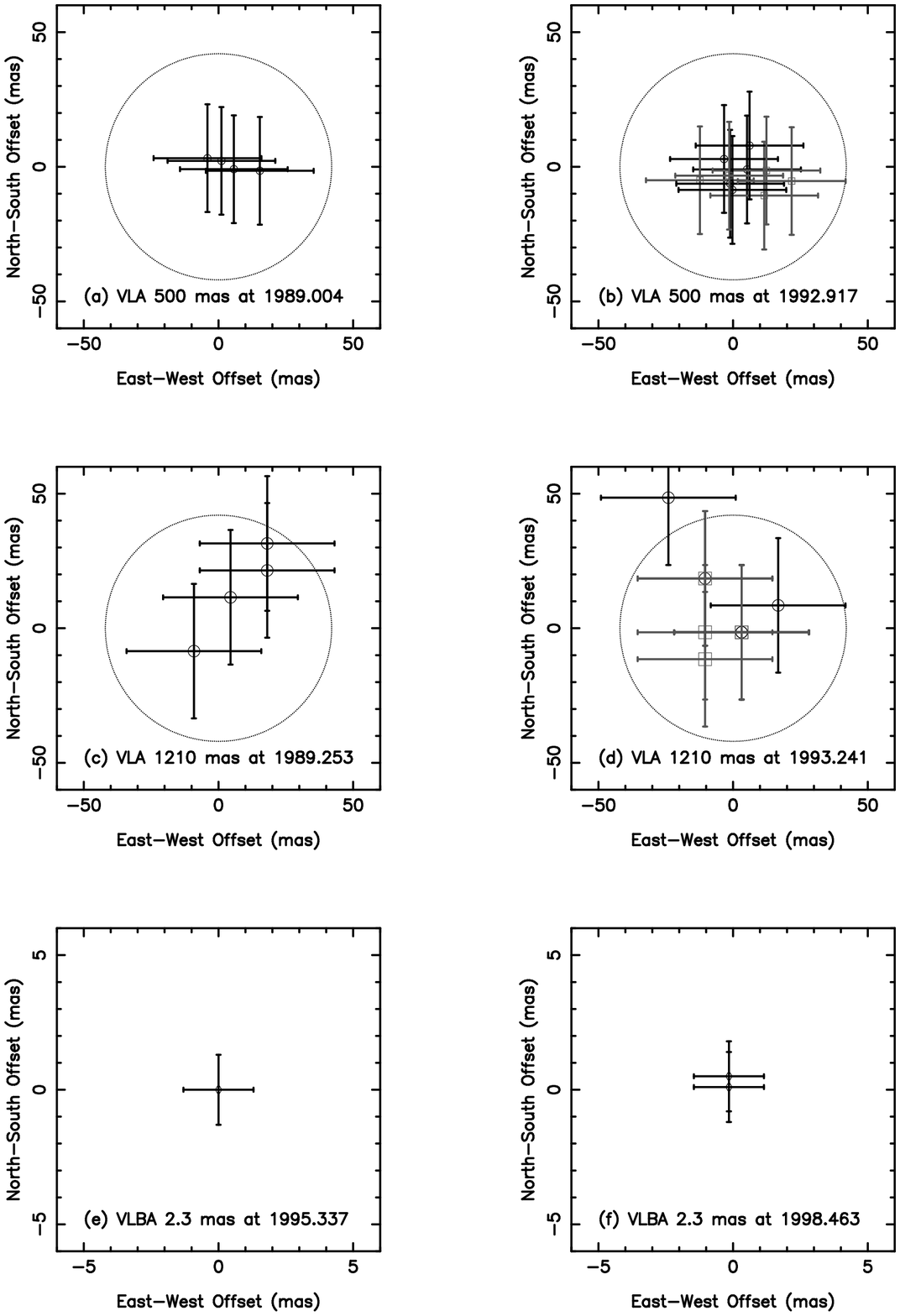]{Astrometric monitoring of the radio continuum
nucleus of NGC\,5548 at the matched angular resolutions and mean
epochs indicated.  Data are from Table~1.  Symbols have the same
meaning as in Figure~6.  Section~4.2 describes the significance for
the star cluster models of the circle in panels (a)-(d).
\label{fig9}}

%
\newpage 
   \epsscale{1.4}
   \plotone{f1.eps}

   \centerline{Figure~1}

\newpage 
   \epsscale{1.0}
   \plotone{f2-xxx.eps}

   \centerline{Figure~2}

 \newpage 
   \epsscale{1.0}
   \plotone{f3-xxx.eps}

   \centerline{Figure~3}

\newpage 
   \epsscale{1.0}
   \plotone{f4.eps}

   \centerline{Figure~4}

\newpage 
   \epsscale{1.4}
   \plotone{f5.eps}

   \centerline{Figure~5}

\newpage 
   \epsscale{0.8}
   \plotone{f6.eps}

   \centerline{Figure~6}

\newpage 
   \epsscale{0.8}
   \plotone{f7.eps}

   \centerline{Figure~7}

\newpage 
   \epsscale{0.4}
   \plotone{f8.eps}

   \centerline{Figure~8}

\newpage 
   \epsscale{0.8}
   \plotone{f9.eps}

   \centerline{Figure~9}

\clearpage

\begin{deluxetable}{lccccllcc}
\footnotesize 
\tablecaption{Photometric and Astrometric Monitoring \label{tab1}} 
\tablewidth{0pc} 
\tablehead{ \colhead{UT}         & 
            \colhead{ }          & 
            \colhead{Resolution} & 
            \colhead{Freq.}      & 
            \colhead{$\sigma_I$} & 
            \colhead{Peak}       & 
            \colhead{Position}   & 
            \colhead{$\sigma_A$} & 
            \colhead{Flux}       \nl 
            \colhead{Epoch}      & 
            \colhead{Array}      & 
            \colhead{(mas)}      & 
            \colhead{(GHz)}      & 
            \colhead{($\mu$Jy~ba$^{-1}$)}  &
            \colhead{$\alpha_{\rm J2000}$} & 
            \colhead{$\delta_{\rm J2000}$} & 
            \colhead{(mas)}      & 
            \colhead{Density}    } 
\startdata 
1992.814 ...& VLA &  500& 8.4&  75& 
$14^h 17^m 59\fs5401$& $+25\arcdeg 08\arcmin 12\farcs604$& 20 &$2.61\pm0.11$\tablenotemark{a}\nl
1988.913 ...& VLA &  500& 8.4&  43&  14 17 59.5409& +25 08 12.603& 20& 1.84$\pm$0.07\tablenotemark{a}\nl
1988.989 ...& VLA &  500& 8.4&  44&  14 17 59.5398& +25 08 12.607& 20& 1.97$\pm$0.07\tablenotemark{a}\nl
1989.036 ...& VLA &  500& 8.4&  36&  14 17 59.5394& +25 08 12.608& 20& 2.05$\pm$0.07\tablenotemark{a}\nl
1989.077 ...& VLA &  500& 8.4&  44&  14 17 59.5402& +25 08 12.604& 20& 1.90$\pm$0.07\tablenotemark{a}\nl
1989.184 ...& VLA & 1210& 8.4&  37&  14 17 59.5411& +25 08 12.626& 25& 2.28$\pm$0.08\tablenotemark{a}\nl
1989.230 ...& VLA & 1210& 8.4&  33&  14 17 59.5401& +25 08 12.616& 25& 2.51$\pm$0.08\tablenotemark{a}\nl
1989.279 ...& VLA & 1210& 8.4&  38&  14 17 59.5391& +25 08 12.596& 25& 2.55$\pm$0.09\tablenotemark{a}\nl
1989.318 ...& VLA & 1210& 8.4&  32&  14 17 59.5411& +25 08 12.636& 25& 2.14$\pm$0.07\tablenotemark{a}\nl
1992.814 ...& VLA &  500& 8.4&  67&  14 17 59.5401& +25 08 12.604& 20& 2.61$\pm$0.10\tablenotemark{a}\nl
1992.814 ...& VLA &  500& 4.9&  76&  14 17 59.5406& +25 08 12.594& 20& 2.94$\pm$0.12\tablenotemark{a}\nl
1992.858 ...& VLA &  500& 8.4&  66&  14 17 59.5395& +25 08 12.608& 20& 2.74$\pm$0.11\tablenotemark{a}\nl
1992.858 ...& VLA &  500& 4.9&  96&  14 17 59.5388& +25 08 12.600& 20& 3.03$\pm$0.13\tablenotemark{a}\nl
1992.891 ...& VLA &  500& 8.4&  56&  14 17 59.5402& +25 08 12.613& 20& 2.67$\pm$0.10\tablenotemark{a}\nl
1992.891 ...& VLA &  500& 4.9& 105&  14 17 59.5406& +25 08 12.603& 20& 2.95$\pm$0.14\tablenotemark{a}\nl
1992.997 ...& VLA &  500& 8.4&  50&  14 17 59.5396& +25 08 12.598& 20& 2.72$\pm$0.10\tablenotemark{a}\nl
1992.997 ...& VLA &  500& 4.9&  78&  14 17 59.5413& +25 08 12.600& 20& 3.07$\pm$0.12\tablenotemark{a}\nl
1993.025 ...& VLA &  500& 8.4&  69&  14 17 59.5397& +25 08 12.596& 20& 2.79$\pm$0.11\tablenotemark{a}\nl
1993.025 ...& VLA &  500& 4.9& 111&  14 17 59.5396& +25 08 12.602& 20& 3.08$\pm$0.14\tablenotemark{a}\nl
1993.162 ...& VLA & 1210& 8.4&  49&  14 17 59.5400& +25 08 12.603& 25& 4.28$\pm$0.14\tablenotemark{a}\nl
1993.162 ...& VLA & 1210& 4.9&  64&  14 17 59.5400& +25 08 12.603& 25& 3.74$\pm$0.13\tablenotemark{a}\nl
1993.203 ...& VLA & 1210& 8.4&  47&  14 17 59.5380& +25 08 12.653& 25& 4.29$\pm$0.14\tablenotemark{a}\nl
1993.203 ...& VLA & 1210& 4.9&  68&  14 17 59.5390& +25 08 12.593& 25& 3.88$\pm$0.14\tablenotemark{a}\nl
1993.249 ...& VLA & 1210& 8.4&  47&  14 17 59.5410& +25 08 12.613& 25& 4.84$\pm$0.15\tablenotemark{a}\nl
1993.249 ...& VLA & 1210& 4.9&  65&  14 17 59.5390& +25 08 12.603& 25& 4.00$\pm$0.14\tablenotemark{a}\nl
1993.277 ...& VLA & 1210& 8.4&  43&  14 17 59.5400& +25 08 12.603& 25& 5.26$\pm$0.16\tablenotemark{a}\nl
1993.277 ...& VLA & 1210& 4.9&  64&  14 17 59.5400& +25 08 12.603& 25& 4.38$\pm$0.15\tablenotemark{a}\nl
1993.315 ...& VLA & 1210& 8.4&  46&  14 17 59.5390& +25 08 12.623& 25& 5.72$\pm$0.18\tablenotemark{a}\nl
1993.315 ...& VLA & 1210& 4.9&  62&  14 17 59.5390& +25 08 12.623& 25& 4.60$\pm$0.15\tablenotemark{a}\nl
1995.337 ...& VLBA& 2.3,1.1,$-6\arcdeg$&
              8.4&  92&  14 17 59.53973& +25 08 12.6048& 1.3& 2.40$\pm$0.30\tablenotemark{b}\nl
1998.445 ...& VLBA& 2.3,1.1,$-6\arcdeg$&
              8.4&  55&  14 17 59.53972& +25 08 12.6049& 1.3& 1.70$\pm$0.20\tablenotemark{b}\nl
1998.481 ...& VLBA& 2.3,1.1,$-6\arcdeg$&
              8.4&  58&  14 17 59.53972& +25 08 12.6053& 1.3& 1.70$\pm$0.20\tablenotemark{b}\nl
\enddata
\tablenotetext{a}{Peak in mJy~ba$^{-1}$.}
\tablenotetext{b}{Total in mJy from areal integration.}
\end{deluxetable}

\end{document}